\documentclass[12pt]{article}
\newlength{\pubnumber} 

\catcode`\@=11
\@addtoreset{equation}{section}

\def\section{\@startsection{section}{1}{\z@}{3.5ex plus 1ex minus .2ex}
 {2.3ex plus .2ex}{\large\bf}}
\def\subsection{\@startsection{subsection}{2}{\z@}{2.3ex plus .2ex}
 {2.3ex plus .2ex}{\bf}}

\begin{document}

\begin{titlepage}
\samepage{
\setcounter{page}{0}
\rightline{\tt hep-th/0312216}
\rightline{December 2003}
\vfill
\begin{center}
   {\Large \it  Adventures in Thermal Duality (I): \\}
   {\Large \bf
            Extracting Closed-Form Solutions
         for Finite-Temperature Effective
        Potentials\\ in String Theory\\}
\vfill
   {\large
      Keith R. Dienes\footnote{
     E-mail address:  dienes@physics.arizona.edu}
        $\,$ and $\,$ Michael Lennek\footnote{
     E-mail address:  mlennek@physics.arizona.edu}
    \\}
\vspace{.10in}
 {\it  Department of Physics, University of Arizona, Tucson, AZ  85721  USA\\}
\end{center}
\vfill
\begin{abstract}
  {\rm  
   Thermal duality, which relates the physics of closed strings at
   temperature $T$ to the physics at the inverse temperature $1/T$,
   is one of the most intriguing features of string thermodynamics. 
   Unfortunately, the classical definitions of thermodynamic quantities
   such as entropy and specific heat are not invariant under
   the thermal duality symmetry.
   In this paper, we investigate whether there might nevertheless
   exist special
   solutions for the string effective potential 
   such that the duality symmetry will be preserved for {\it all}\/ 
   thermodynamic quantities.  
   Imposing this as a constraint, we derive 
   a series of unique functional forms for the complete 
   temperature-dependence of the required string effective potentials.
   Moreover, we demonstrate that these solutions successfully 
   capture the leading temperature behavior of a variety 
   of actual one-loop
   effective potentials for duality-covariant finite-temperature string 
   ground states.
   This leads us to conjecture that 
   our solutions might actually represent the {\it exact}\/ effective potentials 
   when contributions from all orders of perturbation theory are included.  
  }
\end{abstract}
\vfill
\smallskip}
\end{titlepage}

\setcounter{footnote}{0}

\def\beq{\begin{equation}}
\def\eeq{\end{equation}}
\def\beqn{\begin{eqnarray}}
\def\eeqn{\end{eqnarray}}
\def\half{{\textstyle{1\over 2}}}
\def\etainv{{\overline{\eta}}}
\def\calO{{\cal O}}
\def\calE{{\cal E}}
\def\calT{{\cal T}}
\def\calM{{\cal M}}
\def\calF{{\cal F}}
\def\calY{{\cal Y}}
\def\calV{{\cal V}}
\def\rep#1{{\bf {#1}}}
\def\ie{{\it i.e.}\/}
\def\eg{{\it e.g.}\/}

\newcommand{\newc}{\newcommand}
\newc{\gsim}{\lower.7ex\hbox{$\;\stackrel{\textstyle>}{\sim}\;$}}
\newc{\lsim}{\lower.7ex\hbox{$\;\stackrel{\textstyle<}{\sim}\;$}}

\hyphenation{su-per-sym-met-ric non-su-per-sym-met-ric}
\hyphenation{space-time-super-sym-met-ric}
\hyphenation{mod-u-lar mod-u-lar--in-var-i-ant}


\def\inbar{\,\vrule height1.5ex width.4pt depth0pt}

\def\IC{\relax\hbox{$\inbar\kern-.3em{\rm C}$}}
\def\IQ{\relax\hbox{$\inbar\kern-.3em{\rm Q}$}}
\def\IR{\relax{\rm I\kern-.18em R}}
 \font\cmss=cmss10 \font\cmsss=cmss10 at 7pt
\def\IZ{\relax\ifmmode\mathchoice
 {\hbox{\cmss Z\kern-.4em Z}}{\hbox{\cmss Z\kern-.4em Z}}
 {\lower.9pt\hbox{\cmsss Z\kern-.4em Z}}
 {\lower1.2pt\hbox{\cmsss Z\kern-.4em Z}}\else{\cmss Z\kern-.4em Z}\fi}

\long\def\@caption#1[#2]#3{\par\addcontentsline{\csname
  ext@#1\endcsname}{#1}{\protect\numberline{\csname
  the#1\endcsname}{\ignorespaces #2}}\begingroup
    \small
    \@parboxrestore
    \@makecaption{\csname fnum@#1\endcsname}{\ignorespaces #3}\par
  \endgroup}
\catcode`@=12

\input epsf

\section{Introduction}
\setcounter{footnote}{0}

Some of the most intriguing features of string theory have been
the existence of numerous dualities which connect physics in
what would otherwise appear to be vastly dissimilar regimes.
Such dualities include strong/weak coupling duality (S-duality)
as well as large/small compactification radius duality (T-duality),
and together these form the bedrock upon which much of our
understanding of the full, non-perturbative moduli space
of string theory is based.

There is, however, an additional duality which has received far
less scrutiny:  this is {\it thermal duality}\/, which relates
string theory at temperature $T$ with string theory at the inverse temperature
$T_c^2/T$ where $T_c$ is a critical (or self-dual) temperature related
to the string scale.  Thermal duality follows naturally from T-duality
and Lorentz invariance, and thus has roots which are as deep
as the dualities that occur at zero temperature.
Given the importance of dualities of all sorts in
extending our understanding of the unique features of
non-perturbative string theory,
we are led to ask what new insights can be gleaned from
a study of thermal duality.

In this paper, we shall focus on the first feature that
immediately strikes any student of this subject:  classical
thermodynamics, as currently formulated, is not invariant
(or covariant) under thermal duality.
While certain thermodynamic quantities such as the free energy
and the internal energy of an ideal closed string gas exhibit
invariances (or covariances) under thermal duality transformations,
other quantities such as entropy and specific heat do not.

In this paper, we shall investigate whether 
thermal duality might nevertheless happen to be preserved
for special choices of the effective potential.
In other words, we shall investigate whether it is possible 
to construct an effective potential such that {\it all}\/ 
corresponding physically relevant thermodynamic quantities
will turn out to be duality covariant.
Thus, in this way, we seek to exploit thermal duality
in order to constrain the effective potential in a manner that
transcends a direct order-by-order perturbative calculation.

Remarkably, we shall find that 
there exist a unique series of functional forms which
have this property.
Moreover, we shall demonstrate that these solutions
successfully capture the leading temperature dependence of 
the one-loop effective potentials
for a variety of finite-temperature string ground states involving
time/temperature compactifications on $S^1$ (circles)
and $S^1/Z_2$ (orbifolds) in all dimensions $D\geq 2$.
The precision with which this occurs
leads us to conjecture that our solutions
might actually represent the {\it exact}\/ solutions for the
corresponding string effective potentials when 
results from all orders of perturbation theory are included.

Note that a preliminary summary of some of these results has appeared in
Ref.~\cite{summary}.  Our goal here is to provide a more complete and self-contained
discussion and derivation of these results.
There are, however, numerous topics pertaining to string thermodynamics
which we will not address in this paper.  
These include the nature of the Hagedorn phase transition
as well as the Jeans instability and general issues concerning
the interplay between gravity and thermodynamics.
It would be interesting to explore the extent to 
our results concerning thermal duality can shed light on these issues, 
and we hope to address
these questions in future work.

\section{Thermal duality and the rules of thermodynamics}
\setcounter{footnote}{0}

Let us begin by quickly presenting some of the key ideas
that will be relevant for our discussion.
Our goal will be to highlight the manner in which 
the rules of standard thermodynamics 
generally tend to break 
thermal duality.

Just as in ordinary statistical mechanics, the fundamental
quantity of interest in string thermodynamics is 
the one-loop thermal string partition function $Z_{\rm string}(\tau,T)$.
This partition function generally exhibits the symmetries 
of the underlying theory.
For example,
we shall assume that $Z_{\rm string}$ is invariant under modular transformations:
\beq
       Z_{\rm string}(\tau+1,T) ~=~ Z_{\rm string}(-1/\tau,T) ~=~ Z_{\rm string}(\tau,T)~,
\eeq
where $\tau$ is the complex modular parameter describing the shape of the
one-loop (toroidal) worldsheet.
Modular invariance is required for the consistency of the corresponding
closed string model, and arises from the assumption of conformal invariance
at the one-loop level.

More importantly, however, we shall 
also assume that $Z_{\rm string}$ is invariant under thermal duality: 
\beq
       Z_{\rm string} (\tau,T_c^2/T)~ =~ Z_{\rm string}(\tau,T)~
\eeq
where $T_c$ is the self-dual temperature.
Thermal duality also has deep roots (for early papers, see
Refs.~\cite{OBrienTan,AlvOsoNPB304,AtickWitten,AlvOsoPRD40,OsoIJMP,Polbook}).
In general, finite-temperature effects can be
incorporated into string theory~\cite{Pol86}
by compactifying an additional time dimension on a circle (or orbifold~\cite{shyam1})
of radius $R_T = (2\pi T)^{-1}$.
However, Lorentz invariance guarantees that the properties of this
extra time dimension should be the same as those of the original space
dimensions, and T-duality~\cite{SakaiSenda,Nairetal,Sathiapalan}
 tells us that closed string theory
on a compactified space dimension of radius $R$ is indistinguishable
from that on a space of radius $R_c^2/R$ where $R_c= \sqrt{\alpha'}$ is the
self-dual radius.  Together, these symmetries then imply thermal duality,
with $T_c\equiv M_{\rm string}/2\pi$. 
Note that the thermal duality symmetry
holds to all orders in perturbation theory~\cite{AlvOsoPRD40}.

All thermodynamic quantities of interest are generated
from $Z_{\rm string}$.
The finite-temperature vacuum amplitude $\calV(T)$ is given by~\cite{Pol86,McClainRoth,OBrienTan}
\beq
    {\cal V}(T) ~\equiv~ -\half \,{\cal M}^{D-1}\, \int_{\cal F} {d^2 \tau\over ({\rm Im} \,\tau)^2}
             Z_{\rm string}(\tau,T)
\label{Vdef}
\eeq
where ${\cal M}\equiv M_{\rm string}/2\pi$ is the reduced string scale;
$D$ is the number of non-compact spacetime dimensions;
and ${\cal F}\equiv \lbrace \tau:  |{\rm Re}\,\tau|\leq \half,
 {\rm Im}\,\tau>0, |\tau|\geq 1\rbrace$ is the fundamental domain
of the modular group.
Note that $T_c = {\cal M}$.
In general, $\calV(T)$ plays the role usually taken by the logarithm of the
statistical-mechanical partition function.  Because of its role in
governing the dynamics of the theory, we shall occasionally refer to 
the vacuum amplitude $\calV(T)$
as the ``effective potential'' even though this terminology is often
used instead to describe the free energy $F$.  
Given this definition for $\calV$, the free energy $F$, internal
energy $U$, entropy $S$, and specific heat $c_V$ then follow
from the standard thermodynamic definitions:
\beq
          F = T \calV~,~~~~
         U = - T^2 {d\over dT} \calV~,~~~~
         S = -{d\over dT} F~,~~~~ c_V = {d\over dT} U~.
\label{usualrelations}
\eeq

It is easy to see that the thermal duality invariance
of $Z_{\rm string}$ is inherited by some of its descendants.
Since $\calV$ is just the modular integral of $Z_{\rm string}$, 
$\calV$ is also invariant under thermal duality transformations:
\beq
         \calV(T_c^2/T) ~=~ \calV(T)~.
\eeq
Likewise, it is easy to verify that the free energy $F$ and the internal
energy $U$ transform {\it covariantly}\/ under thermal duality:
\beq
          F(T_c^2/T) = \left( {T_c\over T}\right)^2 F(T)~,~~~~~~
          U(T_c^2/T) = -\left( {T_c\over T}\right)^2 U(T)~.
\label{FUtrans}
\eeq
Thus, these quantities also respect the thermal duality symmetry;
in fact, this symmetry sets a zero for the internal energy such that
$U(T_c)=0$.

Unfortunately, the entropy and specific heat fail  
to have any closed transformation properties under the thermal duality
symmetry.  
Specifically, we find 
\beqn
      S(T_c^2/T) &=& -S(T) - 2 F(T)/T~,\nonumber\\
      c_V(T_c^2/T) &=&  c_V(T) - 2U(T)/T~.
\label{failure}
\eeqn 
This failure to transform covariantly suggests that entropy and specific
heat are improperly defined from a string-theoretic
standpoint.  At best, they are not the proper ``eigenquantities''
which should correspond to physical observables.

It is easy to diagnose the source of this problem.
In general, a function $f(T)$ will be called thermal duality covariant
with weight $k$ and sign $\gamma= \pm 1$ if, under the thermal
duality transformation $T\to T_c^2/T$, we find
\beq
         f(T) ~\to~ f(T_c^2/T) ~=~  \gamma\, (T_c/T)^k\, f(T)~.
\label{covariant}
\eeq
Thus, $\calV$ has $(k,\gamma)=(0,1)$,
while $F$ and $U$ have $(k,\gamma)=(2,1)$ and $(2,-1)$ respectively.
Note that $\gamma=\pm 1$ are the only two possible choices consistent
with the $\IZ_2$ nature of the thermal duality transformation.
In general, multiplication by $T$ is a covariant
operation, resulting in a function with weight $k+2$ and the same
sign for $\gamma$.
However, the temperature derivative $d/dT$ generally breaks
duality covariance.  To see this, let us imagine that $f(T)$ has
weight $k$ and sign $\gamma$.   Evaluating $df/dT$ at temperature
$T_c^2/T$, we then find
\beqn
   \left\lbrack {df\over dT}\right\rbrack (T_c^2/T)
   &=& {d\over d(T_c^2/T)} f(T_c^2/T)\nonumber\\
   &=& - \gamma \left( {T\over T_c}\right)^2 {d\over dT} \left[ (T_c/T)^k f(T)\right]\nonumber\\
   &=&  - \gamma \left({ T_c\over T}\right)^{k-2} \left(
                     {df\over dT} - {k f\over T}\right) ~.
\label{deriv}
\eeqn
Thus, as a result of the second term above, we see that
$df/dT$ fails to transform covariantly under the thermal duality transformation
unless $f$ itself has $k=0$.
Since the vacuum amplitude $\calV$ has $k=0$,
this explains why the internal energy $U$ continues to be duality covariant
(with $k=2$) even though it involves a temperature derivative.  However,
since the free energy $F$ and the internal energy $U$ each
already have $k=2$, we see that subsequent derivatives yield quantities (such as
the entropy $S$ and specific heat $c_V$) which are no longer duality covariant.

\section{Special solutions for string effective potentials}
\setcounter{footnote}{0}

Let us now consider whether there might exist special
finite-temperature vacuum amplitudes $\calV(T)$
in which thermal
duality covariance is preserved for all thermodynamic quantities.
In other words, we shall seek special solutions for $\calV(T)$ such that
 {\it all}\/ of its thermodynamic descendants turn out to be duality
covariant,
even though the rules by which these quantities are calculated
explicitly break this symmetry.
We emphasize that in choosing this line of attack, we are necessarily 
losing generality;  we are essentially limiting our attention 
to special, highly symmetric string ground states.
Nevertheless, as we shall see, 
it is important to investigate this possibility.

\subsection{General approach}

In order to proceed along these lines,
we first need to address a general mathematical question:
from amongst all duality-covariant functions $f(T)$ of weight $k$
and sign $\gamma$, are there any {\it special}\/ functions $f(T)$ 
for which $df/dT$ ``accidentally'' turns out to be covariant?

Given the derivative in Eq.~(\ref{deriv}),
we see that there is only one way in which $df/dT$
can possibly be thermal duality covariant:
we must have
\beq
       {df\over dT} ~-~ {\it k f(T) \over T} ~=~ -\delta  \left( {T_c\over T}\right)^\ell {df\over dT}
\label{constraint}
\eeq
for some sign $\delta$  and exponent $\ell$.
If Eq.~(\ref{constraint}) is satisfied, then 
we see from Eq.~(\ref{deriv}) that
$df/dT$ will indeed be covariant, with sign $\gamma \delta$  and weight $k+\ell-2$.
Note that we must have $\delta= \pm 1$ in order to produce a consistent 
sign for $df/dT$.  (The minus sign in front of $\delta$
has been inserted for future convenience.)

It is not difficult to find solutions for $f(T)$ in Eq.~(\ref{constraint}), since
this is nothing but a linear first-order differential equation.
For $\ell \not= 0$, 
we thus obtain the general solution
\beq
              f ~\sim~   (T^\ell + \delta   T_c^\ell )^{k/\ell} ~
\label{solution}
\eeq
where we are disregarding an overall, arbitrary, 
$T$-independent normalization factor.
However, in this derivation we assumed that 
$f$ has weight $k$ and sign $\gamma$.  Checking the solution in Eq.~(\ref{solution}),
we find that this does not restrict the value of $\ell$, but does require that
$ \delta^{ k/\ell} = \gamma$.

By contrast, if $\ell=0$ in Eq.~(\ref{constraint}), we obtain a non-zero solution
for $f(T)$ only if $\gamma=1$ and $\delta= +1$:
\beq
             f~\sim~ T^{k/2}~.
\label{solution2}
\eeq
As required, this also has weight $k$.

Thus, from amongst all possible covariant functions $f(T)$ with weight $k$ and sign $\gamma$, 
we have found that only an extremely restrictive form for $f(T)$ guarantees that $df/dT$ is
also thermal duality covariant:
either $f(T)$ must have the form given in Eq.~(\ref{solution})  
where $\ell\not= 0$ is arbitrary and 
where $\delta^{k/\ell}=\gamma$, with $\delta = \pm 1$;
or $f(T)$ must have the form given in Eq.~(\ref{solution2}), which 
can occur only if $\gamma=1$.  
Of course, overall multiplicative factors of $T_c$ can always be introduced
in either expression as needed on dimensional grounds.

\subsection{Preserving duality covariance for entropy and specific heat:\\
   A thermal duality ``bootstrap''} 

Using this, let us now reconsider our original thermodynamic problem.
We begin with a vacuum amplitude $\calV$, which we assume invariant
under thermal duality transformations.  Thus, $\calV$ necessarily has $k=0$
and $\gamma=1$.  From this, we proceed to derive $F$ and $U$.
Once again, these quantities are also automatically duality covariant;
they each have weight $k=2$ and their signs are
$\gamma = +1$ and $-1$ respectively.
Up to this point, the functional forms for $\calV$, $F$, and $U$ are completely
arbitrary (subject to the above constraints on their weights and signs).
However, it is in calculating $S$ and $c_V$
that potential difficulties arise, for we must demand
that $S$ and $c_V$ be simultaneously covariant as well.
This then provides two new non-trivial constraints
on the forms of $F$ and $U$, as discussed above.
Working backwards, this then provides a very restrictive set 
of possibilities for the vacuum amplitudes $\calV$ from which
both $F$ and $U$ are derived.
In other words, we will have essentially used a ``bootstrap'' 
formed by demanding the covariance of $S$ and $c_V$ to deduce 
a particular form (or set of forms) of the vacuum amplitude $\calV$.

Carrying out this calculation is relatively straightforward.
We first focus on the entropy $S$.
In order for $S$ to be thermal duality covariant, the free energy 
$F$ (which must have weight $k=2$ and sign $\gamma = 1$) is required
to take the form
\beq
           F(T) ~\sim~ -{ (T^\ell + \delta T_c^\ell)^{2/\ell}\over T_c }~
\label{Fform}
\eeq
where 
\beq
           \delta^{2/\ell} ~=~ 1~.
\label{deltacond}
\eeq
Note that the factor of $T_c$ in the denominator of Eq.~(\ref{Fform})
has been inserted on dimensional grounds (where we implicitly express our
thermodynamic quantities in units of ${\cal M}^{D-1}$); 
likewise, we have also inserted an overall minus 
sign for future convenience.
Also note that Eq.~(\ref{deltacond}) restricts us to $\delta= +1$ for even $\ell$, 
but allows $\delta= \pm 1$ for odd $\ell$.
This form for $F$ guarantees that $S$, which takes the form
\beq
           S(T) ~\sim~  2 \,{T^{\ell -1}\over T_c} \, (T^\ell + \delta T_c^\ell)^{2/\ell -1}~,
\eeq
is covariant with weight $\ell$ and sign $\delta$.

We are of course deliberately disregarding the $\ell=0$ possibility, stemming
from Eq.~(\ref{solution2}), that $F(T)\sim T$.  We reject this possibility
not only because this would make $F(T)$ independent of $T_c$ 
(which is unexpected from a string calculation), but also because 
it leads to an entropy which is completely
temperature-independent and hence unphysical. 

Given $F(T)$ in Eq.~(\ref{Fform}), we immediately determine that $\calV(T)$ must
take the general form
\beq
          \calV(T) ~\sim~ -{ (T^\ell + \delta T_c^\ell)^{2/\ell}\over T T_c }~.
\label{Vform}
\eeq
Note that this is indeed invariant under thermal duality transformations, as required.
This in turn implies that $U(T)$ must have the general form
\beq
          U(T) ~\sim~  {1\over T_c}\,  (T^\ell + \delta T_c^\ell)^{2/\ell-1}
           \, (T^\ell - \delta T_c^\ell)~,
\label{Uform}
\eeq
which is of course consistent with our requirement that $U$ have weight $2$ and
sign $-1$.
Thus, up to this point, we have found that the entropy 
will be thermal duality covariant (along with the effective potential, the free
energy, and the internal energy)
if and only if $\calV(T)$ takes the form (\ref{Vform}).

We now impose our requirement that $c_V$ also be thermal duality covariant.
As we shall see, this will provide a constraint on the value of $\ell$.
Since $U(T)$ is given in Eq.~(\ref{Uform}),
we can immediately calculate the specific heat, obtaining
\beq
           c_V(T)~\sim~  2 \,{T^{\ell-1}\over T_c}\,
                   (T^\ell +\delta T_c^\ell)^{2/\ell -2} \,
         \left\lbrack   T^\ell + (\ell -1)\delta T_c^\ell \right\rbrack~.
\label{cVform}
\eeq
Clearly, this quantity fails to be duality covariant unless the
final factor in square brackets takes the form $T^\ell \pm \delta T_c^\ell$
with $\delta=\pm 1$, or unless this factor takes the form
$T^\ell$ (in which case this factor
joins with the overall $T^{\ell-1}$ prefactor to modify the duality weight
of $c_V$).
These two options occur only for $\ell=2$ or $\ell=1$ respectively.

Note that the $\ell=1,2$ cases provide maximal duality symmetry for our 
solutions.  Indeed, in these cases,
our solution for $U(T)$ also simultaneously takes the form
\beq
              U(T) ~\sim~   {(T^{m} + \epsilon   T_c^{m} )^{2/m} \over T_c}
\label{Uform2}
\eeq
for some $m$ and sign $\epsilon= \pm 1$ (with $\epsilon^{2/m} = -1$),
as required from Eq.~(\ref{solution})
in order to yield a covariant specific heat $c_V=dU/dT$.
Moreover, since the specific heat is also given by the relation
$c_V=T dS/dT$, our solution for $S(T)$ also takes this same special
form in these cases. 

\begin{figure}
  ~\vskip -0.53 truein
\centerline{
   \epsfxsize 2.7 truein \epsfbox {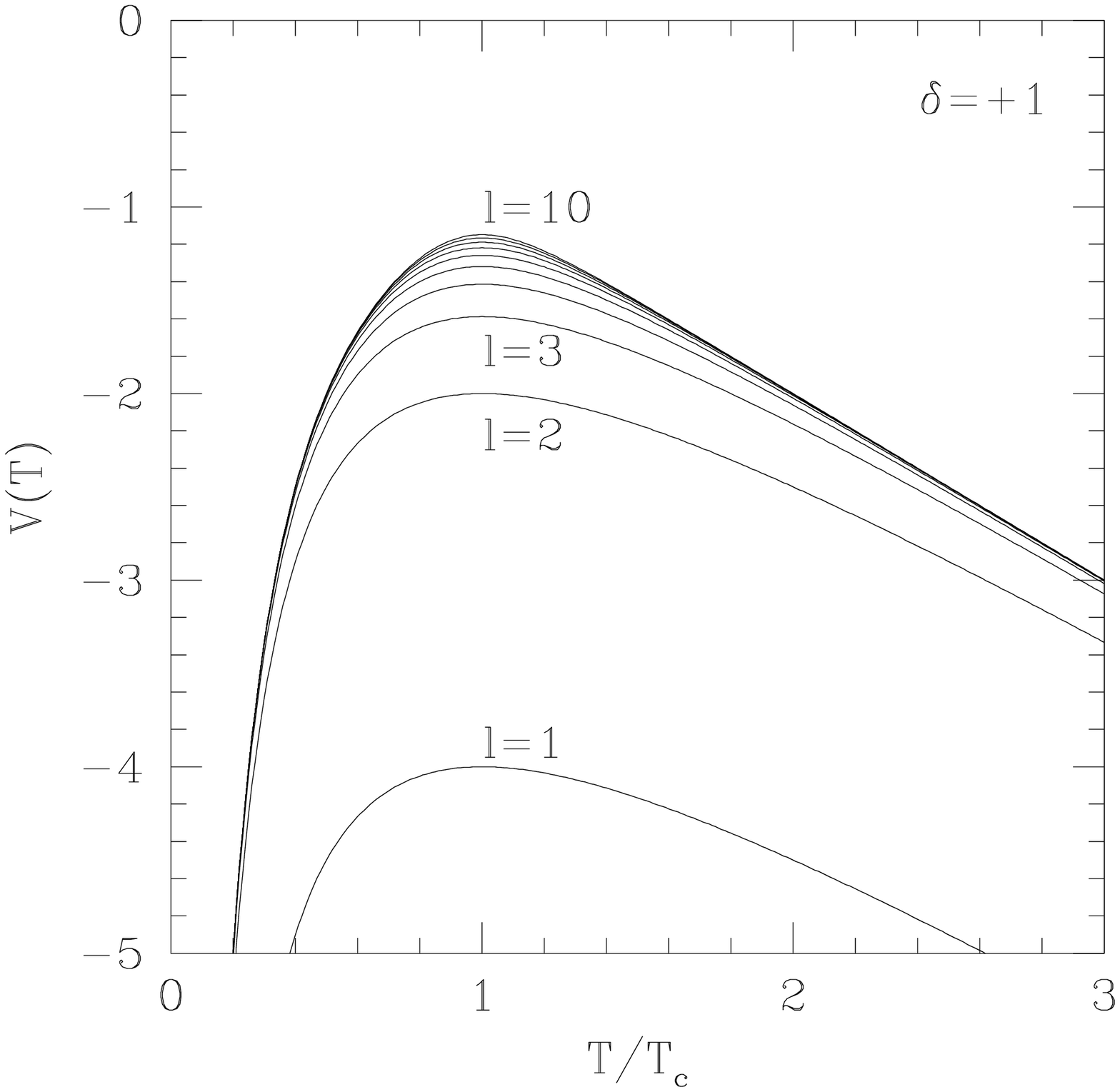}
    }
\centerline{
   \epsfxsize 2.7 truein \epsfbox {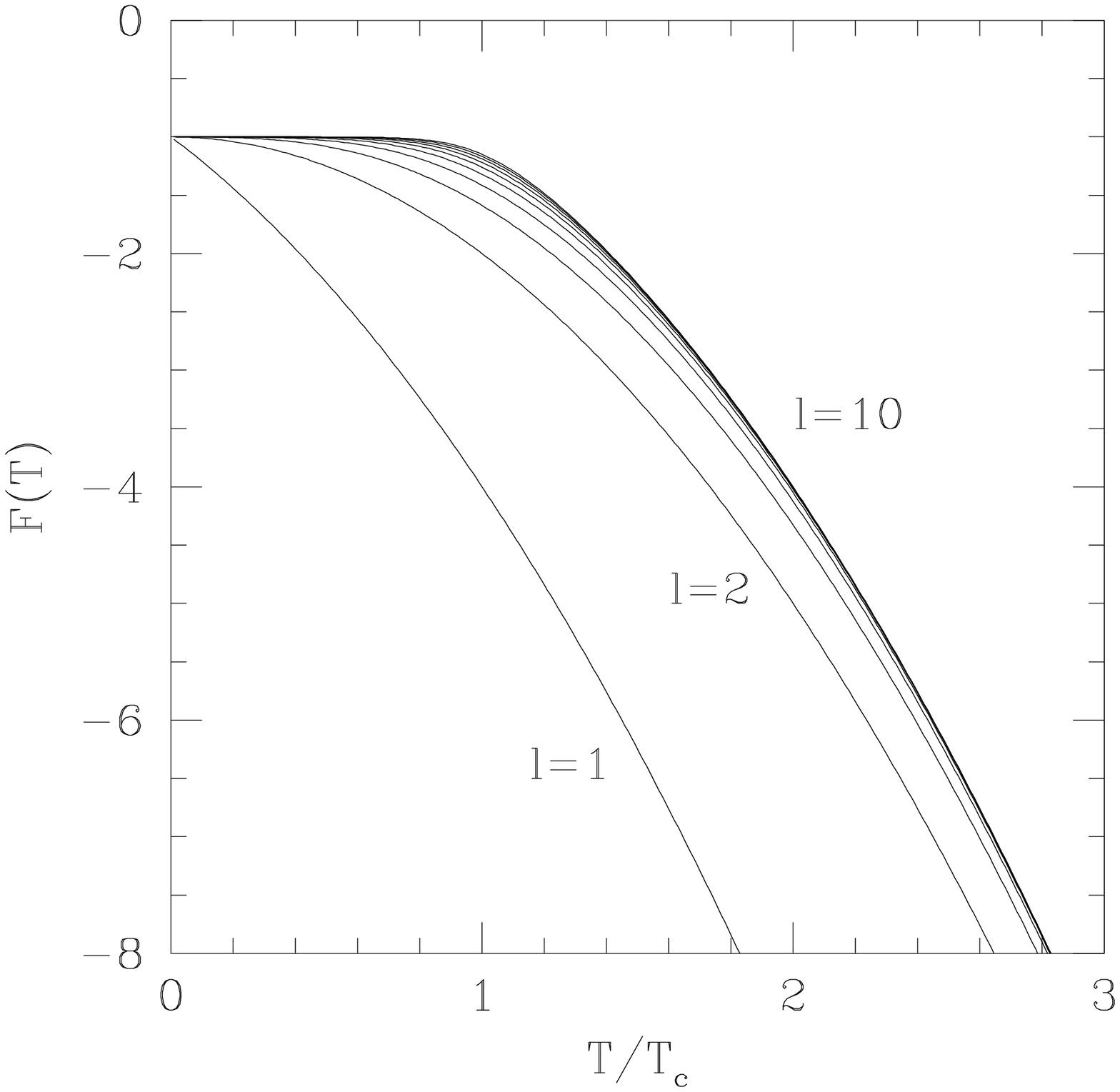}
   \epsfxsize 2.7  truein \epsfbox {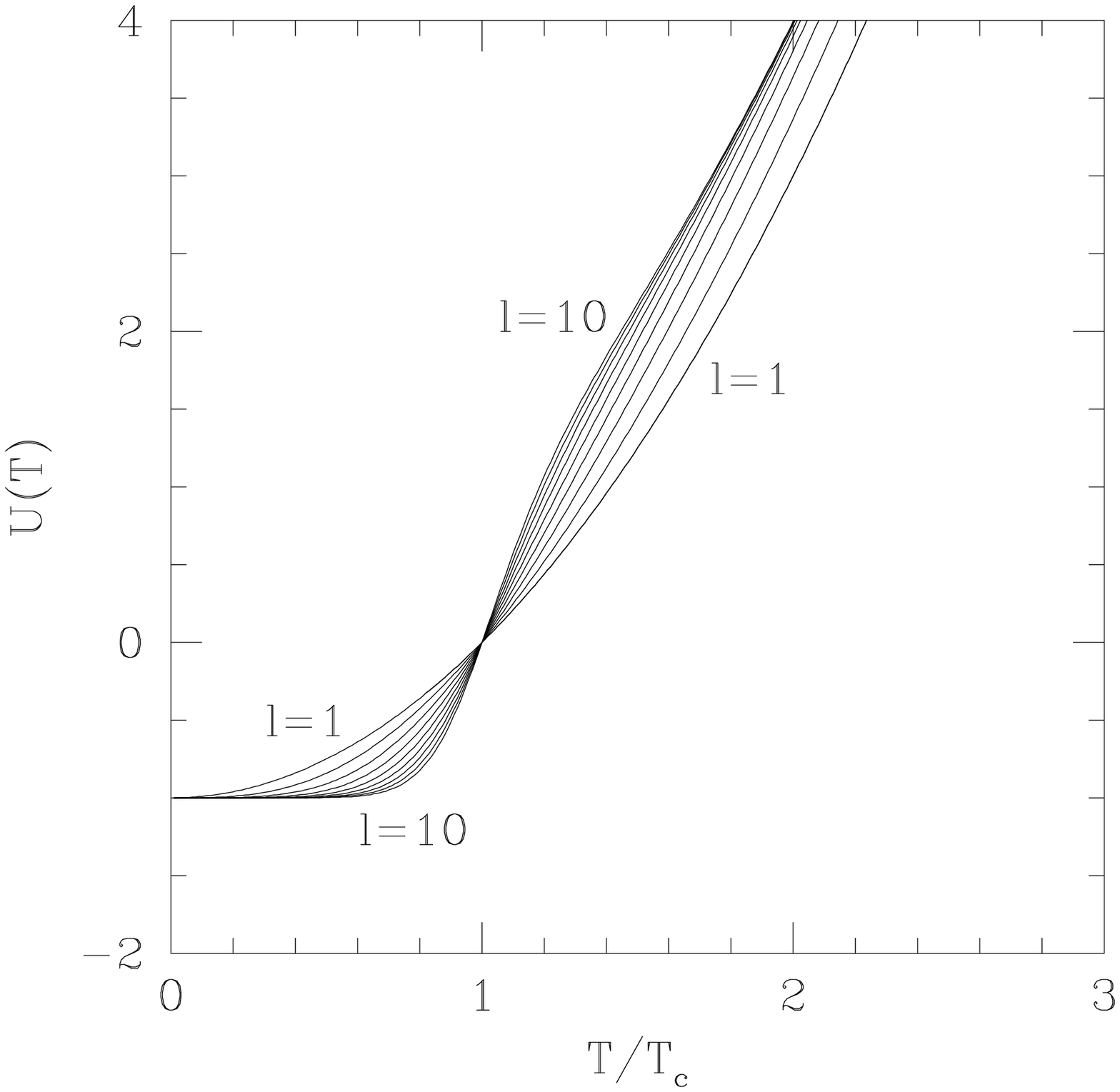}
    }
\centerline{
   \epsfxsize 2.7  truein \epsfbox {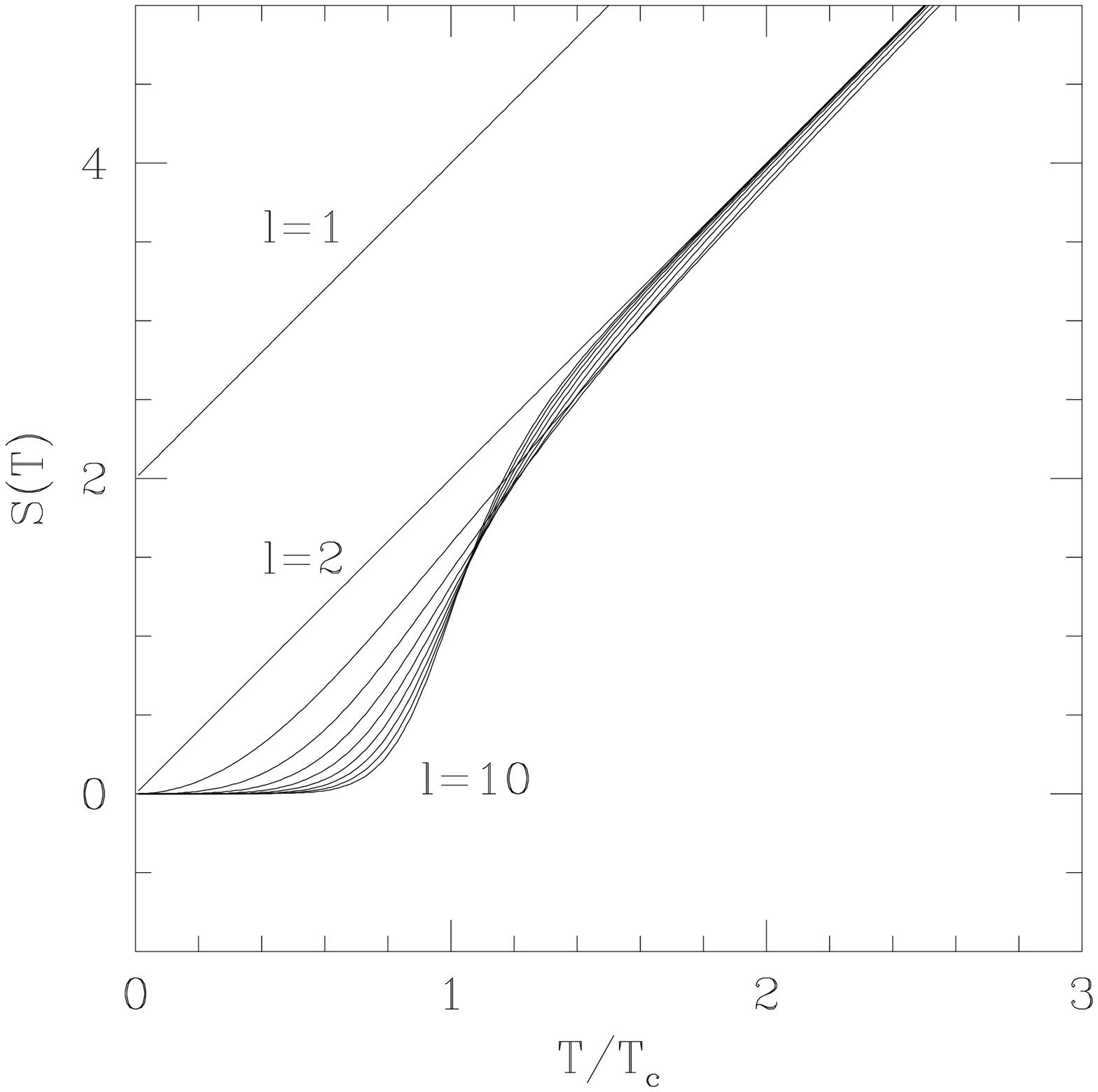}
   \epsfxsize 2.7  truein \epsfbox {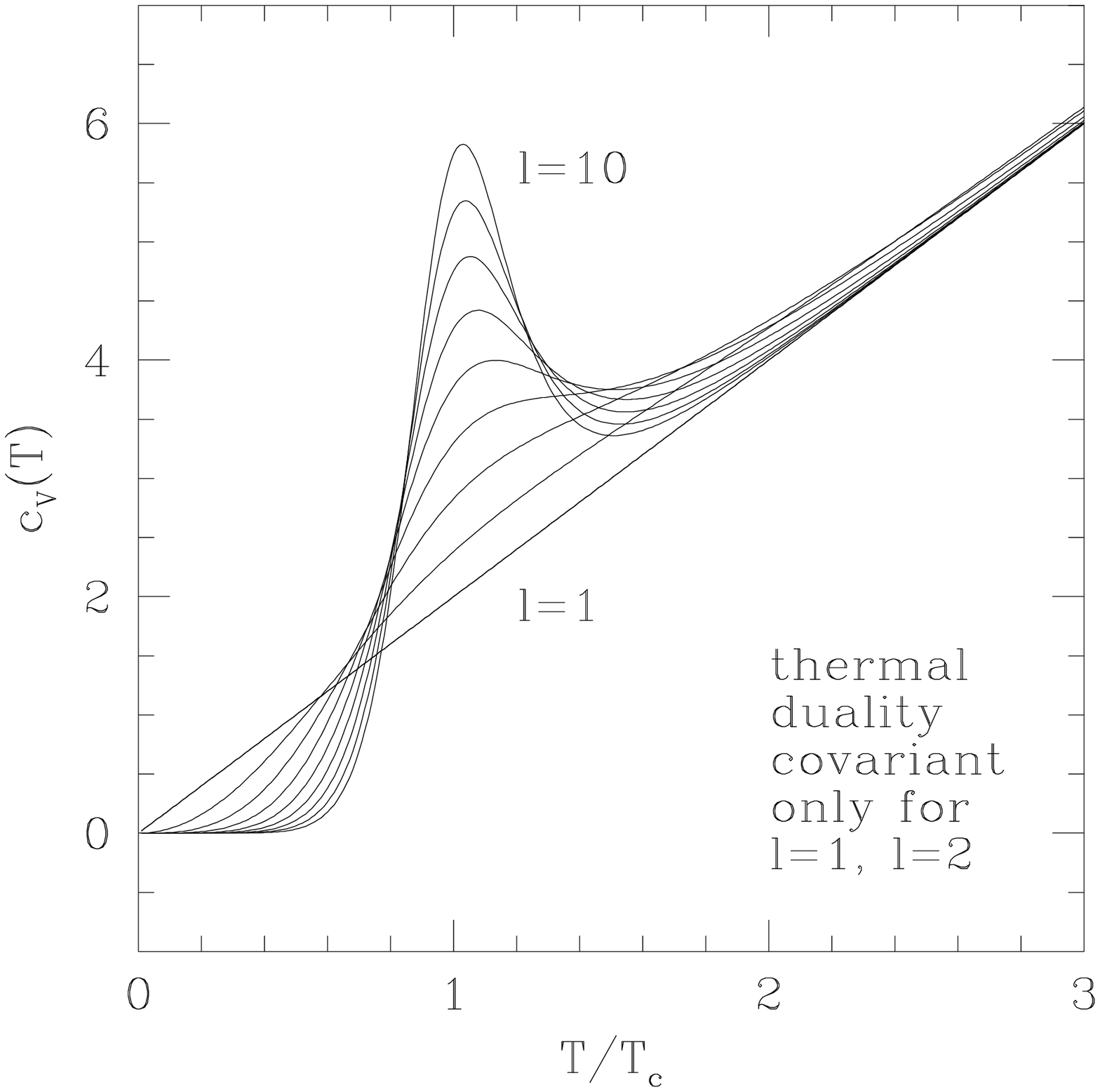}
   }
\caption{The thermodynamic quantities $\calV$, $F$, $U$, $S$, and $c_V$
   in Eq.~(\protect\ref{lsoln}),
   plotted as functions of $T$ for $1\leq \ell\leq 10$ and $\delta=+1$,
    in units of ${\cal M}\equiv M_{\rm string}/2\pi= T_c$. 
   All quantities except for $c_V$ are thermal duality covariant for all $\ell$,
   while  $c_V$ is covariant only for $\ell=1,2$.  For these values of $\ell$, 
   the entropy and specific heat are exactly linear functions of $T$.
    Note that $c_V$ develops a discontinuity as $\ell\to \infty$,
    suggesting the emergence of a second-order phase transition in this limit.}
\label{fig1}
\end{figure}

Thus, summarizing, we see that our requirement of preserving
general covariance for our thermodynamic quantities 
forces them to have a particular form:
\beqn
          \calV^{(\ell)}(T) &\sim& - (T^\ell + \delta T_c^\ell)^{2/\ell}/ T T_c \nonumber\\
          F^{(\ell)}(T) &\sim& - (T^\ell + \delta T_c^\ell)^{2/\ell}/ T_c \nonumber\\
          U^{(\ell)}(T) &\sim& \phantom{-} (T^\ell + \delta T_c^\ell)^{2/\ell-1}
                           (T^\ell - \delta T_c^\ell)/ T_c \nonumber\\
          S^{(\ell)}(T) &\sim&  2 \,T^{\ell -1} (T^\ell + \delta T_c^\ell)^{2/\ell -1}/ T_c\nonumber\\
        c^{(\ell)}_V(T) &\sim&  2 \,T^{\ell-1} (T^\ell +\delta T_c^\ell)^{2/\ell -2}
         \left\lbrack   T^\ell + (\ell -1)\delta T_c^\ell \right\rbrack/ T_c~
\label{lsoln}
\eeqn
where 
\beq
               \delta ~=~ \cases{   +1 &  $\ell$ even\cr
                                 \pm 1 &  $\ell$ odd~.\cr}
\label{deltacases}
\eeq
These solutions are plotted in Fig.~\ref{fig1},
and ensure that $\calV$, $F$, $U$, and $S$ are all thermal
duality covariant for any value of $\ell$.  
While $\calV$, $F$, and $U$ have duality weights 
$(k,\gamma)= (0,1)$, 
$(2,1)$, and
$(2,-1)$ respectively, 
the entropy $S$ has duality weight and sign $(k,\gamma)=(\ell,\delta)$.
Observe that the traditional relation $U=F+TS$ continues to hold for all $\ell$.

However, $c_V$
will also be thermal duality covariant if and only if $\ell=1$ or $\ell=2$.
The explicit solutions in these cases reduce to
\beqn
        \ell=2:~~~~~~\calV^{(2)}(T) &=&   - (T^2 + T_c^2)/(T T_c)\nonumber\\
        F^{(2)}(T) &=&   - (T^2 + T_c^2)/T_c  \nonumber\\
        U^{(2)}(T) &=&   \phantom{-} (T^2 - T_c^2)/ T_c \nonumber\\
        S^{(2)}(T) &=&   \,\,2 T/T_c\nonumber\\
      c_V^{(2)}(T) &=&   \,\,2 T/T_c ~    
\label{l2soln}
\eeqn
and
\beqn
        \ell=1:~~~~~~\calV^{(1)}(T) &=&   - (T +\delta T_c)^2/(T T_c) \nonumber\\
        F^{(1)}(T) &=&   - (T +\delta T_c)^2/T_c \nonumber\\
        U^{(1)}(T) &=&   \phantom{-} (T^2 - T_c^2)/ T_c \nonumber\\
        S^{(1)}(T) &=&   \,\,2 (T/T_c + \delta)\nonumber\\
      c^{(1)}_V(T) &=&   \,\,2 T/T_c ~    
\label{l1soln}
\eeqn
where $\delta = \pm 1$.
Note that $c_V$ has weight $k_c=2$ and
sign $\gamma_c=1$  for both the $\ell=2$ and $\ell=1$ solutions.

Clearly, the $\ell=1$ and $\ell=2$ solutions are closely related.
They share the same expressions for $U$ and $c_V$,
yet their expressions for $\calV$, $F$, and $S$
are shifted by constants or extra linear terms:
\beqn
       \calV^{(\ell=1)} &=&  \calV^{(\ell=2)}  ~-~ 2\delta~\nonumber\\
       F^{(\ell=1)} &=&  F^{(\ell=2)}  ~-~ 2\delta T~\nonumber\\
       S^{(\ell=1)} &=&  S^{(\ell=2)}  ~+~ 2\delta~.
\label{shifts}
\eeqn
This shift symmetry will be important in the following.

These $\ell=1,2$ solutions also exhibit other intriguing symmetries.  For example,
since $F(T)\sim -(T^2+T_c^2)/T_c$ and
$U(T)\sim (T^2-T_c^2)/T_c$ for $\ell=2$,
we see that $F(iT)=U(T)$ and $U(iT)=F(T)$.
In other words, we have the formal symmetry
\beq
          T~\to~ iT:~~~~~~~~~ F ~\longleftrightarrow ~U~.
\label{FUsymmetry}
\eeq
Since $F=T\calV$ and $U=-T^2 d\calV/dT$, this immediately leads to a symmetry
for $\calV(T)$:
\beq
            iT\, \calV(iT) ~=~ -T^2 {d\calV\over T}~,
\eeq
or equivalently
\beq
              {d\calV\over dT} ~=~ {\calV(iT)\over iT}~.
\label{Vsymm}
\eeq
This symmetry is remarkable because it relates the troublesome
temperature derivative $d\calV/dT$ to $\calV$ itself.
Since $\calV(T)$ is defined through a modular integral
as in Eq.~(\ref{Vdef}), 
this implies that
quantities involving the temperature
derivative of $\calV$ can now be written as
\beq
          {d\calV\over dT} ~=~   -\half \,{\cal M}^{D-1}\, {1\over iT}\,
            \int_{\cal F} {d^2 \tau\over ({\rm Im} \,\tau)^2}
             Z_{\rm string}(iT)~.
\eeq
Moreover, it is easy to show that just as the symmetry
(\ref{FUsymmetry}) leads to the symmetry (\ref{Vsymm}),
it also leads to a symmetry for the second derivative:
\beq
    {d^2 \calV\over dT^2} ~=~ {\calV(T)\over T^2} - {1\over T} {d\calV\over dT} ~=~
             {1\over T^2} \left\lbrack \calV(T) + i \calV(iT) \right\rbrack~.
\label{Vsymm2}
\eeq
It is, in fact,  this identity that enforces $S=c_V$ for our $\ell=2$ solutions.
Similar symmetries also hold for the $\ell> 2$ solutions.

\section{Comparison with explicit one-loop calculations:\\
     Temperature dependence of effective potentials
}
\setcounter{footnote}{0}

We now seek to determine the extent to which 
our closed-form solutions match the results 
of explicit one-loop modular integrations
of the sort that
can emerge from actual finite-temperature string ground states.
Such comparisons are extremely important 
because our derivation of the functional forms
given in Sect.~3 was ``top-down'', based entirely
on thermal duality symmetries, and did not make
reference to any perturbative, order-by-order calculation.
Moreover, our discussion was completely model-independent.

Nevertheless, as we shall now discuss, our expressions successfully 
capture the leading temperature dependence of 
the one-loop effective potentials
for a variety of 
modular integrals involving 
time/temperature compactifications on $S^1$ (circles)
and $S^1/Z_2$ (orbifolds).  Moreover, this will occur for
all spacetime dimensions $D\geq 2$.
As we shall see, the precision with which this occurs
will ultimately lead us to conjecture that our solutions
actually represent the {\it exact}\/ solutions for the
corresponding string effective potentials when 
results from all orders of perturbation theory (and perhaps
even non-perturbative effects) are included.

\subsection{Calculating the one-loop effective potential}

Let us first recall the calculation of the one-loop 
effective potential
for a finite-temperature string ground state  
in which the time/temperature direction is compactified 
on a circle.  This is appropriate, \eg, for compactifications
of the bosonic string, and we shall consider such circle compactifications
for most of what follows.
In $D$ spacetime dimensions, the
one-loop effective potential for such compactifications takes the form
in Eq.~(\ref{Vdef}), where
\beq
         Z_{\rm string}(\tau,T) ~\equiv~
         Z_{\rm model}(\tau) \, Z_{\rm circ}(\tau,T)~.
\label{parfunct}
\eeq
Here $Z_{\rm model}$ represents the trace over the Fock space
of states (\ie, the partition function) of the 
string model in question, formulated at zero temperature.
For example, 
in the case of the bosonic string compactified
to $D$ spacetime dimensions, $Z_{\rm model}$ takes the
general form  
\beq
         Z_{\rm model} ~=~ ({\rm Im}\,\tau)^{1-D/2}\,
          \, {\overline{\Theta}^{26-D} \Theta^{26-D}
         \over \overline{\eta}^{24} \eta^{24} }
\label{Zstringgenform}
\eeq
where the numerator
$ \overline{\Theta}^{26-D} \Theta^{26-D}$
schematically represents a sum over the $2(26-D)$-dimensional
compactification lattice for left- and right-movers. 
Note that in general, $Z_{\rm model}$ is the quantity which appears
in the calculation of the one-loop cosmological
constant (vacuum energy density) of the model:
\beq
         \Lambda ~\equiv~ 
     -\half \,{\cal M}^D\, \int_{\cal F} {d^2 \tau\over ({\rm Im} \,\tau)^2}
             Z_{\rm model}~.
\label{Lambdadef}
\eeq
By contrast, the remaining factor $Z_{\rm string}$
represents the sum over Matsubara frequencies.
For extended objects such as strings, this includes not only 
``momentum'' Matsubara states but also ``winding'' Matsubara states. 
For time/temperature circle compactifications, 
$Z_{\rm string}$ is given by\footnote{
  Since we are defining $Z_{\rm circ}$ to represent the
  sum over Matsubara frequencies, we do not include
  the Dedekind $\eta$-function denominators which 
  would traditionally be required in order to interpret 
  $Z_{\rm circ}$ as the partition function of a boson 
  compactified on a circle of radius 
  $R_T\equiv (2\pi T)^{-1}$. 
  This does not represent a violation of modular
  invariance, since the extra factor of  
  $\sqrt{ {\rm Im}\, \tau}$ in 
  Eq.~(\ref{Zcircdef}) compensates for their absence.
  Note that this factor offsets the similar factors
  in $Z_{\rm model}$ (just as the summation in $Z_{\rm circ}$
  combines with the lattice sums in $Z_{\rm model}$),
   thereby effectively
  reducing by one the dimensionality of the resulting
  finite-temperature string model compared  
  with the dimensionality of the original string 
  model at zero temperature.}
\beq
     Z_{\rm circ}(\tau,T)~=~ 
     \sqrt{ {\rm Im}\, \tau}\,
    \sum_{m,n\in\IZ} \, 
      \overline{q}^{(ma-n/a)^2/4}  \,q^{(ma+n/a)^2/4}
\label{Zcircdef}
\eeq
Here the double sum tallies both the Matsubara momentum
and winding states, with $q\equiv \exp(2\pi i\tau)$ and 
$a\equiv 2\pi T/M_{\rm string} = T/T_c$ where
$T_c\equiv M_{\rm string}/2\pi={\cal M}$.
Thus, thermal duality symmetry is nothing but the symmetry
($a\leftrightarrow 1/a, m\leftrightarrow n$) in Eq.~(\ref{Zcircdef}).  

It is important to emphasize that a factorization of
the form given in Eq.~(\ref{parfunct}) holds only for the
simplest finite-temperature string constructions (such
as for the bosonic string).  In more realistic setups,
simple factorizations such as this are not possible,
and one typically has more complicated configurations 
(see, \eg, Refs.~\cite{Rohm,AlvOsoNPB304,McGuigan,AtickWitten,KounnasRostand}).
In this section, however, we shall confine our attention to this
simplest case because it is the situation in which 
thermal duality is most directly manifest.

\subsection{Asymptotic behavior for low and high temperatures}

Given the form of these partition functions, it is straightforward
to deduce the leading behavior in the $T\to 0$ and $T\to \infty$ limits,
and
verify that this behavior matches the corresponding behavior of 
our solutions in Sect.~3.
Taking the $T\to 0$ limit 
of $Z_{\rm circ}$, we find
\beq
          Z_{\rm circ} ~\to~ {1\over a}~~~~~~~~
    {\rm as}~~ a\to 0~.
\label{Zcirclimit}
\eeq
This implies the limiting behavior
\beq
        \calV(T) ~\sim ~ {\Lambda\over T}~~~~~~~~ {\rm as}~~ T/T_c\to 0~,
\label{Vleadingform}
\eeq 
where $\Lambda$ is the one-loop cosmological constant in Eq.~(\ref{Lambdadef}).
This in turn implies that $F(T)\to \Lambda$ as $T/T_c\to 0$.

However, this leading behavior for $\calV(T)$ and $F(T)$ coincides exactly with the 
$T\to 0$ temperature dependence 
of the solutions found in Sect.~3 for arbitrary $\ell$.  
In fact, this agreement allows us to go one step further and
deduce the overall normalization 
of our solutions for arbitrary $\ell$ with $\delta= +1$: 
\beq
         \calV^{(\ell)}(T) ~=~ {\Lambda \over T_c} \, { (T^\ell + T_c^\ell)^{2/\ell}\over T T_c}~.
\label{Vnormalized}
\eeq

We can also consider the opposite, high-temperature limit 
$T\to \infty$ in Eq.~(\ref{Vnormalized}), obtaining~\cite{AtickWitten,OsoIJMP,Polbook}
\beq
    \calV^{(\ell)}(T)~\sim~ {\Lambda \over T_c} \, {T\over T_c}~~~~~~~~  {\rm as}~~ T\to\infty~.
\eeq
This implies that $F^{(\ell)}(T)\sim T^2$ as $T\to\infty$, correctly reproducing
the celebrated high-temperature behavior which signals the reduced
number of degrees of freedom in finite-temperature string theory 
relative to field theory~\cite{AtickWitten}.
Note that these correct limiting behaviors are obtained for all values of
$\ell$.  

Having thus verified that our solutions $\calV^{(\ell)}(T)$ in 
Sect.~3 correctly reproduce the expected, leading $T\to 0$ and 
$T\to \infty$ behaviors for all $\ell$,
we now turn to a more detailed study of this scaling behavior
as a function of temperature.
It turns out that
this will enable us to understand the role played by the free parameter $\ell$.

In ordinary quantum field theory, the free energy $F(T)$ at large
temperatures typically scales like $T^{D}$ where $D$ is the spacetime 
dimension.  This in turn implies that the entropy $S$ should scale like $T^{D-1}$.
However, as already noted above, in string theory we have $F(T)\sim T^2$ as $T\to \infty$,
implying that $S(T)\sim T$ as $T\to \infty$.
Thus, string theory behaves asymptotically as though it has
an effective dimensionality $D_{\rm eff}=2$.

Of course, the field-theory limit of string theory is expected
to occur for $T\ll T_c$. 
Given this, it is interesting to examine the effective dimensionality
(\ie, the effective scaling exponent)
of our solutions as a function of temperature.
In general, it is easiest to define this
effective dimensionality $D_{\rm eff}(T)$ by considering the entropy:
since $S^{(\ell)}(T)$ is a monotonically increasing function of $T$,
we can define $D_{\rm eff}(T)$ as the effective scaling exponent 
at temperature $T$, setting
$S^{(\ell)}(T)\sim   T^{D_{\rm eff}-1}$. 
We thus have, as a general definition, 
\beq
    D_{\rm eff} ~\equiv~  1+ { d \ln S\over d\ln T} ~=~ 
               1 +  {T\over S} {dS\over dT} ~=~ 1 + {c_V\over S}~,
\label{Deff}
\eeq  
where the last equality follows from the thermodynamic
identity $c_V= T dS/dT$.

\begin{figure}[ht]
\centerline{
   \epsfxsize 3.0 truein \epsfbox {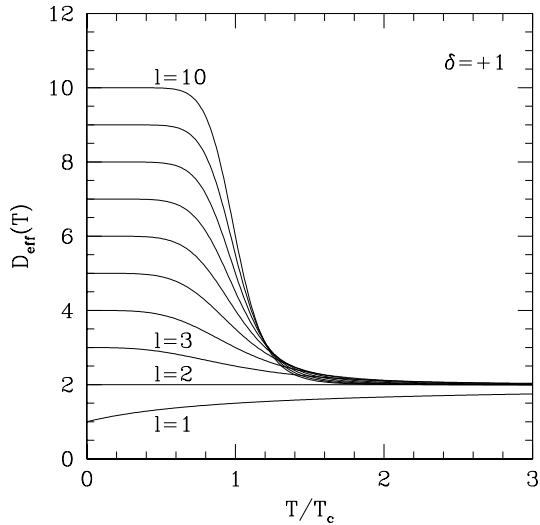}
    }
\caption{The effective dimensionalities $D_{\rm eff}$ 
    of our thermodynamic solutions,
    plotted as functions of $T$
   for $1\leq \ell\leq 10$ and $\delta=+1$.
   All of our solutions successfully 
   interpolate between
   $D_{\rm eff}=\ell$ for $T\ll T_c$ 
   and $D_{\rm eff}=2$ for $T\gg T_c$.
   Only the $\ell=2$ solution has $D_{\rm eff}=2$ for
   all $T$.}
\label{fig2}
\end{figure}

These results for $D_{\rm eff}(T)$ are plotted in Fig.~\ref{fig2}.
As we see, each of our solutions successfully 
interpolates between $D_{\rm eff}=\ell$ for $T\ll T_c$ 
and $D_{\rm eff}=2$ for $T\gg T_c$.
Indeed, only the $\ell=2$ solution has $D_{\rm eff}=2$ for
all $T$.

Given this observation, 
it is now possible to interpret our solutions physically.
For small temperatures $T\ll T_c$, the entropy 
behaves as we expect on the basis of field theory, growing according
to the power-law $S^{(\ell)}(T)\sim T^{\ell-1}$.
Indeed, this is nothing but the high-temperature limit of the
low-energy effective {\it field}\/ theory,
which leads us to interpret $\ell$ as the spacetime dimension $D$.
However, as $T$ approaches the
reduced string scale $T_c$,
we see that this asymptotic behavior begins to change,
with the $T^{\ell-1}$ growth in the entropy ultimately turning
into the expected {\it linear} growth for $T\gg T_c$.
This is then the  asymptotic {\it string}\/ limit.

Of course, our identification of $\ell$ as the spacetime dimension
$D$ is subject to one important caveat.
Since $D$ can be defined only through
the {\it high}\/-temperature limit of the underlying {\it field}\/
theory, our identification of $\ell$ with $D$ assumes that
we can properly identify the {\it high}\/-temperature field-theory
limit with the {\it low}\/-temperature string-theory limit for which
$S^{(\ell)}(T)\sim T^{\ell-1}$.
In other words, this identification of $\ell$ with $D$ is sensitive to the 
manner in which the high-temperature limit of field theory 
matches onto what ultimately becomes the low-temperature 
limit of string theory.  
However, we see from Fig.~\ref{fig2} that in all our solutions,
$D_{\rm eff}$ remains very close to $\ell$ for 
almost all of the temperature range up to $T_c$.
Thus, we expect our association of $\ell$ with $D$ to be reasonably
accurate.  Moreover, 
in the special case with $\ell=2$, we know that
$D_{\rm eff}=2$ for all $T$.  We thus expect 
that this case should correspond to $D=2$ exactly.

If we consider the same issue from the perspective of
the free energy, we can also immediately see the origin 
of this difference between the high-temperature 
scaling behaviors in field theory and in string theory.
Note that our solution for the free energy can be written
as
\beq
      F^{(\ell)}(T) ~=~ \Lambda \left\lbrack  1 + \left( {T\over T_c} \right)^\ell
               \,\right\rbrack^{2/\ell}~
\eeq
where we have inserted the normalization factor $\Lambda$ determined
above.
Expanding this solution for small temperatures, we find
\beq
     F^{(\ell)}(T) ~\sim ~ \Lambda ~+~ {2\Lambda\over \ell} 
               \left( {T\over T_c} \right)^\ell ~+~ ...  
            ~~~~~~~~~ {\rm for}~~ T\ll T_c~.
\label{Flowtemp}
\eeq
Thus, as already observed above, $F^{(\ell)}(T)$ begins with 
a {\it constant term}\/ $\Lambda$;
the field-theoretic power-law scaling $T^\ell$ appears
only at subleading order.  
However, it is precisely this constant term which
ultimately determines
the high-temperature scaling behavior in string theory. 
Recall that if $f$ is a general weight-$k$ covariant function 
scaling as $f(T)\sim T^p$ at small temperatures,
then $f$ must scale
as $f(T)\sim T^{k-p}$ at high temperatures.
Thus, the unusual string-theoretic scaling behavior 
$F(T)\sim T^2$ 
at high temperatures 
can ultimately be attributed to the 
fact that $F(T)$ leads with a constant term
$\Lambda$ at small temperatures.

Many of these facts are already well known as general statements
in the string literature (see, \eg, Ref.~\cite{Polbook}).  
What we are observing here, however, is that our
functional forms correctly exhibit all of these properties simultaneously.

\subsection{Direct comparison for all temperatures}

Since we have already determined that our solutions exhibit the expected 
low- and high-temperature scaling behaviors for all $\ell$, 
the question now boils down to whether these solutions correctly match
the expected temperature dependence at {\it intermediate}\/ temperatures
where $T\approx T_c$.   
In other words, we now wish to do a direct 
comparison at all temperatures.

For simplicity, we begin in $D=2$ by considering model-independent situations in which
we set\footnote{
   Setting $Z_{\rm model}=1$ does not violate the form given 
   in Eq.~(\ref{Zstringgenform})
   since we can equivalently write $Z_{\rm model}=
    |\vartheta_2\vartheta_3\vartheta_4 |^{16}/ (2^{16} |\eta|^{48})$
   where $\vartheta_i$ are the Jacobi theta functions
   satisfying $\vartheta_2\vartheta_3\vartheta_4 = 2\eta^3$.}
$Z_{\rm model}$ to 1.  
Since $Z_{\rm model}$ does not contain any temperature dependence of its own,
this simplification enables us to focus directly on the temperature dependence 
arising from $Z_{\rm circ}$.
Our expression for $\calV(T)$ from Eqs.~(\ref{Vdef}),
(\ref{parfunct}), and (\ref{Zcircdef})
then reduces to 
\beq
    \calV^{(D=2)}(T) ~=~ -\half \,{\cal M} \,\int_{\cal F} 
         {d^2 \tau\over ({\rm Im} \,\tau)^{3/2}}
    \,\sum_{m,n\in\IZ} \, 
      \overline{q}^{(ma-n/a)^2/4}  \,q^{(ma+n/a)^2/4}~,
\label{Vd2}
\eeq
with a corresponding ``cosmological constant'' given by
\beq
 \Lambda~=~ -\half\, {\cal M}^2\, \int_{\cal F} 
   {d^2 \tau\over ({\rm Im}\,\tau)^2} ~=~ -{\pi\over 6}\, {\cal M}^2~. 
\eeq

Since $D=2$ in this case, we expect that our expression for $\calV^{(D=2)}(T)$
should directly match onto our $\ell=2$ solution. 
Remarkably, this is exactly what
occurs:  $\calV^{(D=2)}(T)$ 
is {\it exactly equal}\/ to 
our $\ell=2$ solution $\calV^{(\ell=2)}(T)$ with $\delta=+1$:
\beq
    \calV^{(D=2)}(T) ~=~  -{\pi \over 6}\, {T^2 + T_c^2 \over T}~,
\label{miracle}
\eeq
where we have used the fact that $T_c={\cal M}$.
Note that Eq.~(\ref{miracle}) 
holds for {\it all}\/ temperatures $T$.
Thus, our closed-form $\ell=2$ solution from Sect.~3 exactly reproduces
the complete temperature dependence corresponding to the $D=2$ circle
compactification in Eq.~(\ref{Vd2})!  
 
Mathematically, this is a rather surprising result.  
In Eq.~(\ref{Vd2}), the temperature
dependence of $\calV^{(D=2)}(T)$ enters only through the quantity $a\equiv T/T_c$ 
which appears in
the exponents of $q$ and $\overline{q}$;  this temperature dependence, 
taking the form of a sum of $\tau$-dependent exponentials,
is then integrated over the fundamental domain of the modular group.   
Nevertheless, we find that
the net result of this integration is to produce the simple, closed-form result 
given in Eq.~(\ref{miracle}).
Moreover, as we have already seen in Sect.~3, this temperature dependence
is given by precisely the functional form which is necessary in order to ensure
that {\it all}\/ thermodynamic quantities, including {\it both}\/ the entropy and 
the specific heat, are thermal duality covariant.  

This agreement provides an important link between the ``top-down'' analysis of 
Sect.~3 and our direct ``bottom-up'' string calculation.
This agreement is especially illuminating, given that our ``top-down''
derivation made use of a powerful, non-perturbative duality symmetry,
while our ``bottom-up'' string calculation represents only a one-loop result.
Taking this agreement seriously, we are tempted to view 
the one-loop result for this $D=2$ example as ``exact'', receiving
no further contributions at higher loops.  
Of course, in the absence of an actual string model underlying the
expression in Eq.~(\ref{Vd2}), this statement is only meant to be suggestive. 

Before leaving the $D=2$ special case, we remark that
time/temperature circle compactifications 
are not the only possibility in the construction of finite-temperature
string ground states.   
Another choice (perhaps even a preferred choice
phenomenologically~\cite{shyam1}) is to compactify 
on an $S^1/\IZ_2$ orbifold,
\ie, a line segment.
Indeed, under our factorization assumption in Eq.~(\ref{parfunct}),
these two choices represent the only two consistent 
geometries on which a finite-temperature string ground state
may be formulated~\cite{Ginsparg}.
In the case of an orbifold compactification, we simply replace
$Z_{\rm circ}$ in Eq.~(\ref{Zcircdef}) with~\cite{Ginsparg} 
\beq
      Z_{\rm orb}(\tau,T) ~=~    
         \half\, Z_{\rm circ}(\tau,T) ~+~  Z_{\rm circ}(\tau,T_c) ~-~ \half\, Z_{\rm circ}(\tau,T_c/2)~. 
\label{Zorb}
\eeq
In this expression, the first term represents the contributions
from the untwisted states, while the remaining terms are 
 {\it temperature-independent} (\ie, they are evaluated at fixed 
specified temperatures which are independent of $T$) and represent the 
contributions from the twisted states.
Since we already know the complete temperature dependence arising from 
$Z_{\rm circ}$ in Eq.~(\ref{miracle}),  we immediately find that
the effective potential in the orbifold case has the exact closed-form solution
\beq
         \calV_{\rm orb}^{(D=2)} ~=~ -{\pi\over 12} \left\lbrack
                {T^2+T_c^2\over T} ~+~ {3\over 2} \right\rbrack~.
\label{orbsoln}
\eeq

Of course, this is nothing but our $Z_{\rm circ}$ solution,
rescaled and shifted by an additive constant.  However, recall from 
Eq.~(\ref{shifts}) that the $\ell=1$ solution differs from the 
$\ell=2$ solution merely through such an additive shift.
Since the circle solution corresponds to $\ell=2$, this suggests
that our orbifold solution in Eq.~(\ref{orbsoln}) can be expressed
exactly as a linear combination of the $\ell=2$
and $\ell=1$ solutions in Eq.~(\ref{Vnormalized}),
and this is indeed the case:
\beq
         \calV_{\rm orb}^{(D=2)} ~=~  
           {3\over 4} \, \calV^{(\ell=1)} ~+~
           {1\over 4} \, \calV^{(\ell=2)} ~
\label{lincomb}
\eeq
where we have taken $\delta=+1$ in the $\ell=1$ solution.
Once again, we stress that this is an {\it exact}\/ representation
for the complete temperature dependence of the $D=2$ orbifold case.
Note that in writing this expression,
we have identified the normalization constant
$\Lambda = \Lambda_{\rm circ}/2$;  this follows
from the low-temperature limit of Eq.~(\ref{Zorb}), even 
though this $\Lambda$ is no longer the cosmological
constant of the original zero-temperature model.
Also note that even though the $\ell=1,2$ solutions
are shifted relative to each other by an additive constant,
we cannot write $\calV_{\rm orb}$ purely in terms of either of the $\ell=1$
solutions (with $\delta=\pm 1$) 
because the additive shifts in these $\ell=1$ solutions
are $\pm 2$ relative to the $\ell=2$ solution.  
According to Eq.~(\ref{orbsoln}), however,
our shift constant is $3/2$ relative to the $\ell=2$ solution.
This fact has some important consequences which we shall
discuss in Sect.~6.

Given the exact orbifold solution in Eq.~(\ref{orbsoln}),
we can immediately see the thermodynamic effects of 
compactifying the time/temperature dimension on an orbifold 
rather than a circle.  While the internal energy and specific
heat are unaffected by this choice, the free energy picks up
an additional linear term and the entropy picks up an additive
constant.  The latter has been called a ``fixed-point'' 
entropy~\cite{shyam1} since it 
arises from the fixed points of the $S^1/\IZ_2$ orbifold
and  survives even in the $T\to 0$ limit;
in the present case this fixed-point entropy is given exactly as
\beq
          S_{\rm fixed-point} ~=~  \pi/8~. 
\label{fixedpointS}
\eeq

Let us now proceed to consider the case in higher dimensions $D>2$.
As might be expected, things are more complicated.
For arbitrary $D$, the expression in Eq.~(\ref{Vd2}) now generalizes to
\beq
    \calV^{(D)}(T) ~=~ -\half \,{\cal M}^{D-1} \,\int_{\cal F} 
         {d^2 \tau\over ({\rm Im} \,\tau)^{(D+1)/2}}
    \,\sum_{m,n\in\IZ} \, 
      \overline{q}^{(ma-n/a)^2/4}  \,q^{(ma+n/a)^2/4}~,
\label{VDdef}
\eeq
where we incorporate the $D$-dependent prefactor $({\rm Im}\,\tau)^{1-D/2}$
from $Z_{\rm model}$ but continue to disregard the rest of this function
for simplicity. 
While Eq.~(\ref{VDdef}) is not modular invariant, it captures the
dominant $T$- and $D$-dependence that we wish to explore.

As evident from Eq.~(\ref{VDdef}),
the net effect of altering the spacetime dimension is to change the power
of the $({\rm Im}\,\tau)$ factor that appears in the measure of the integral.
If we view the $Z_{\rm circ}$ integrand as a power series in $q$ and $\overline{q}$,
with each term separately integrated and then summed to produce the effective
potential $\calV^{(D)}$, we see that the dominant 
effect of changing the spacetime dimension is to {\it reweight}\/ the contributions 
from each term in the $Z_{\rm circ}$ power series because they are now being integrating 
over the modular-group fundamental domain 
with an altered measure.  
Thus, it is not immediately apparent how the temperature dependence found in
the $\ell=2$ case should change.

Nevertheless, we find that 
our functions $\calV^{(\ell)}(T)$ 
from Sect.~3 continue to successfully capture 
the dominant temperature dependence of the resulting integrals,
with $\ell=D$.
Unlike the case with $D=\ell=2$, this agreement is only approximate
rather than exact.  Nevertheless, we find that this agreement
holds to within one or two percent over the entire temperature 
range $0\leq T \leq \infty$.
Indeed, if we were to superimpose a plot of $\calV^{(D)}(T)$
over the plot of $\calV^{(\ell)}(T)$ in Fig.~\ref{fig1}, 
taking $\ell=D$,
we would not be able to discern the difference at the level of magnification
in Fig.~\ref{fig1}.

Once again, this is a rather striking result, suggesting that our 
functional forms continue to capture the dominant 
temperature dependence, even in higher dimensions.
Of course, for $D>2$, our solutions and the above one-loop results do
not agree exactly.  However, given the significant role 
played by thermal duality in constraining the form of the
effective potential to the specific functional forms that we
have found in Sect.~3, and given the precision with which the
above one-loop results appear to match these functional forms,
it is natural to attribute the failure to obtain an exact
agreement for $D>2$ to the fact that $\calV^{(D)}(T)$ in Eq.~(\ref{VDdef})
is itself only a one-loop approximation.  We thus are led to conjecture that
our functional forms $\calV^{(\ell)}(T)$ indeed represent the exact
solutions for the finite-temperature effective potentials, 
even in higher dimensions,
and that these solutions will emerge only when the 
contributions from all orders in perturbation theory are included.    
Viewed from this perspective, it is perhaps all the more remarkable
that we found an exact agreement for $D=2$, suggesting that
the one-loop result is already exact in this special case,
with no further renormalization.

Let us now consider what happens if 
we do not make the simplification that $Z_{\rm model}=1$
[or $Z_{\rm model}= ({\rm Im}\,\tau)^{1-D/2}$].
Of course, in order to select an appropriate $Z_{\rm model}$, we
must actually construct a bona-fide string model (\eg,
a specific bosonic string compactification);  moreover,
this model must be tachyon-free if our effective potential
is to be finite.
These constraints force $Z_{\rm model}$ to take the form
$Z\sim 1 + \sum_{mn} a_{mn} \overline{q}^m q^n$ 
where $a_{mn}=0$ if $m=n<0$ (no physical tachyons).
The presence of the leading constant term
in the power expansion means that the leading temperature dependence
of $\calV^{(D)}$ will continue to be the same as we had when we merely
set $Z_{\rm model}=1$.
Indeed, the contributions from the higher terms 
in $Z_{\rm model}$ are exponentially suppressed relative
to the leading term, which
means that the net effect of the extra, model-dependent terms 
in $Z_{\rm model}$ is to provide an exponentially
suppressed reweighting of the contributions from the different  
terms in the power-series expansion of $Z_{\rm circ}$.
Thus, the net effect of inserting a non-trivial $Z_{\rm model}$
into $\calV^{(D)}$ is merely to change the {\it subleading}\/ temperature
dependence in a model-dependent way.  Thus, 
we conclude that the leading temperature dependence continues
to be captured by our solutions $\calV^{(\ell)}(T)$ even when
$Z_{\rm model}\not=1$;  indeed, this is the universal,
model-independent contribution.
Moreover, if our conjecture is correct, then 
we expect these subleading model-dependent contributions 
to be washed out as higher-order contributions are included
in the perturbation sum.
Just as for the $D=2$ special case,
similar remarks apply if we replace the thermal compactification
geometry from a circle to an orbifold.

Finally, let us briefly comment on the most general cases
of all, namely those in which the finite-temperature partition
functions do {\it not}\/ factorize as in Eq.~(\ref{parfunct}).
Such cases include compactifications with temperature-dependent
Wilson lines, and are expected to emerge in heterotic or Type~II
theories where non-trivial phases must be introduced 
in the combined thermal partition function
(ultimately due to presence of spacetime fermions). 
For example, non-factorized thermal partition functions 
emerge for finite-temperature string theories whose 
zero-temperature limits are spacetime supersymmetric;
these theories necessarily have thermal partition functions
in which the cancellations inherent in supersymmetry
are non-trivially mixed with the Matsubara 
sums (see, \eg, the 
examples in Refs.~\cite{Rohm,AlvOsoNPB304,McGuigan,AtickWitten,KounnasRostand}).
In such cases, however, the effective potentials do not generally
exhibit thermal duality --- indeed, such theories
may be considered to be 
finite-temperature string ground states in which 
thermal duality is spontaneously broken.
Such theories are therefore beyond the scope of this paper.
We shall, however, present an analysis of such theories 
in Ref.~\cite{III}, where we will show that an analogue of this bootstrap
approach can be developed for such theories as well.

\section{Effective scaling dimensionalities: \\ 
         Connection  to holography?}  
\setcounter{footnote}{0}

In Eq.~(\ref{Deff}),
we defined the notion of an effective dimensionality $D_{\rm eff}$
which governs the scaling behavior of the entropy $S(T)$, with 
$S(T)\sim T^{D_{\rm eff}-1}$.  
As we have seen, 
this scaling coefficient generally ranges from $D_{\rm eff}= D$
as $T\to 0$ to $D_{\rm eff}= 2$ as $T\to \infty$.
The limiting behavior as $T\to 0$ is precisely as expected
on the basis of ordinary quantum field theory,
while the opposite limiting behavior 
as $T\to \infty$ is precisely as required by thermal duality.

This reduction in the effective dimensionality of the system
at high temperatures is extremely reminiscent of 
holography (such an interpretation can also be found, \eg, in Refs.~\cite{shyam1,shyam2}).
Indeed, the scaling of our thermodynamic quantities departs from the
ordinary $D$-dimensional scaling that would be expected on the basis
of quantum field theory, and begins to behave as though
the number of accessible degrees of freedom 
populates not the full $D$-dimensional spacetime,
but rather a subspace of smaller dimensionality.   
Of course, an analysis formulated
in flat space (such as ours) cannot address
questions pertaining to the geometry of this subspace, and thus
cannot determine whether the surviving degrees
of freedom are really to be associated with a subspace or
boundary of the original geometry.  
However, from the restricted perspective emerging from a mere counting
of states, we see our scaling behavior
differs significantly from field-theoretic expectations,
suggesting some sort
of reduction in the effective dimensionality associated
with thermally accessible degrees of freedom as $T\to \infty$.

Of course, taking the $T\to \infty$ limit is merely
of formal interest.  
In a theory with thermal duality, 
there is no difference between the range $T>T_c$ and the range $T<T_c$
since these ranges capture the same physics and are thus indistinguishable.
Or, phrased another way, thermal duality tells us that there is a ``maximum'' 
temperature in the same sense that T-duality tells us there is a minimum radius.
This is also consistent with our expectation that there should
be a Hagedorn-type phase transition at or near $T_c$,
with the theory ultimately entering a new phase marked by new degrees of freedom.
Thus, we should really only consider the range $0\leq T\leq T_c$.
 
Given this, let us consider the value of $D_{\rm eff}$ not
as $T\to 0$ or $T\to \infty$, but as $T\to T_c$. 
As discussed above, this is truly the ``high-temperature'' limit of string theory.
With our specific closed-form solutions $\calV^{(\ell)}(T)$ in Eq.~(\ref{lsoln}), 
the general definition in Eq.~(\ref{Deff}) yields
\beq
    D_{\rm eff}(T) ~=~  {2 T^D + D T_c^D \over T^D + T_c^D }~
\eeq  
where we have identified $\ell=D$.
We thus obtain
\beq
    D_{\rm eff}(T_c)  ~=~  \half \, (2 + D)~.
\eeq

\begin{figure}[ht]
\centerline{
   \epsfxsize 3.0 truein \epsfbox {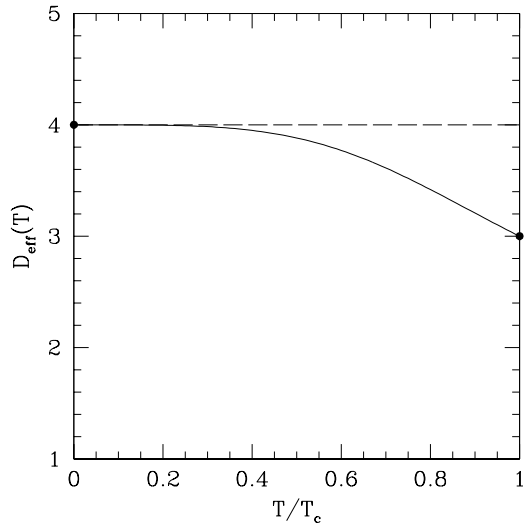}
    }
\caption{The effective dimensionality $D_{\rm eff}$ 
     of our four-dimensional thermodynamic solutions, 
    plotted as a function of $T$.
    These solutions behave ``holographically'' in the 
    range $0\leq T\leq T_c$, with 
     the effective scaling dimensionality  
    falling exactly from $D_{\rm eff}=4$ to $D_{\rm eff}=3$.
    The dotted line indicates the behavior that would be expected
     within quantum field theory.}
\label{holofig}
\end{figure}

This result indicates that 
$D_{\rm eff} (T_c) <D$ for all $D>2$.  In other words, for all $D>2$,
the effective scaling of the number of degrees of freedom 
at high temperatures is reduced compared with our field-theoretic expectations
at low temperatures.
However, taking the predictions of holography seriously, we can ask
when this reduction in $D_{\rm eff}$ is truly 
``holographic'' in the sense that $D_{\rm eff}$ is reduced by exactly 
one unit as $T\to T_c$, dropping from $D$ to $D-1$.  This would be analogous,
for example, to what occurs for black holes, where quantities such as entropy
scale not with the three-volume of the black hole, but with its area.
Remarkably, demanding that $D_{\rm eff}$ drop by precisely one unit yields
\beq
       D_{\rm eff}(T_c) = D-1~~~~~~\Longrightarrow~~~~~~~  D=4~.
\eeq
Thus, we see that it is precisely in four dimensions that our solutions behave
``holographically'' in the range $0\leq T \leq T_c$, with the effective scaling 
dimensionality  falling exactly by one unit from $D_{\rm eff}=4$ 
to $D_{\rm eff}=3$.
This behavior is plotted in Fig.~\ref{holofig}.

While it is tempting to interpret this reduction
in $D_{\rm eff}$ as a holographic effect, 
we again caution that our setup (based on a flat-space calculation)
is incapable of yielding
the additional geometric information that this claim would require.
Such an analysis is beyond the scope of this paper,
and would require reformulating the predictions of thermal
duality for string theories in non-trivial $D$-dimensional backgrounds, 
and then determining whether we could formulate a map between
degrees of freedom in the bulk of the $D$-dimensional volume 
and those on the $(D-1)$-dimensional boundary of this volume. 
Nevertheless, we find this reduction in $D_{\rm eff}$ to be
an extremely intriguing phenomenon, especially since our exact
solutions lead to a strictly ``holographic''
reduction in $D_{\rm eff}$ for $D=4$.
We thus believe that this approach towards understanding
the relation between thermal duality and holography is
worthy of further investigation.

\section{Discussion}
\setcounter{footnote}{0}

In this paper, we set out to address a very simple issue:
even though thermal duality is an apparent fundamental property of string
theory, emerging as a consequence of Lorentz invariance
and T-duality, the rules of classical thermodynamics
do not appear to respect this symmetry.  Even when the 
vacuum amplitude $\calV(T)$ exhibits thermal duality,
thermodynamic quantities such as entropy and
specific heat do not.
Given this situation, we sought to determine
whether special string ground
states might exist
such that thermal duality will nevertheless be
exhibited by all of the usual thermodynamic quantities
of interest.

We began by deriving specific solutions $\calV(T)$ 
such that thermal duality is preserved not only
for the free and internal energies, 
but also for the entropy and specific heat.
The complete set of such solutions is itemized
in Eqs.~(\ref{lsoln}), (\ref{l2soln}), and (\ref{l1soln}).  
While the solutions 
for general $\ell$ preserve thermal duality for all
thermodynamic quantities except the specific heat,
the $\ell=1,2$ solutions preserve thermal duality
for {\it all}\/ of the thermodynamic quantities.

We then investigated the extent to which
these solutions might emerge from modular integrals
of the sort that would be expected in one-loop calculations
from actual string ground states.
Remarkably, we found
that our $\ell=1,2$ closed-form solutions 
provide {\it exact}\/ representations
for $D=2$ modular integrals corresponding to 
time/temperature compactifications on circles and 
orbifolds.  
This agreement 
is  particularly encouraging from a mathematical standpoint,
since our derivation of these functional
forms is entirely ``top-down'', proceeding only from
thermal duality symmetry principles,
and has nothing to do with specific ``bottom-up'' 
constructions involving specific one-loop modular integrations.
The fact that these two approaches agree
exactly, yielding the same results even in highly
simplified cases,
suggests that thermal duality is likely to play an important role
governing self-consistent string ground states.
Indeed, as we saw in Sect.~5, these $\ell=1,2$ solutions 
also ensure that
modular invariance is also preserved for 
all relevant thermodynamic quantities.

By contrast, our remaining $\ell>2$ closed-form solutions 
do not serve as exact representations of 
appropriate $D>2$ modular integrals. 
Nevertheless, we found
that they provide extremely accurate {\it approximations}\/ 
to such integrals in a wide variety of cases.
This led us to conjecture that our $\ell>2$ functional 
forms may indeed provide {\it exact}\/ solutions for
the effective potentials corresponding to wide classes
of finite-temperature string ground states once the contributions
from all orders of perturbation theory (and perhaps even non-perturbative
effects) are included.  After all, our method of deriving these
solutions rests solely on the requirement of thermal duality,
a symmetry which (like the T-duality from which it is derived)
holds to all orders in perturbation theory, and even non-perturbatively.
Thus, if this conjecture is correct, 
it is perhaps not surprising that our $\ell>2$ solutions 
transcend the results of intrinsically one-loop calculations.

In this connection, it is important to stress that 
the free energy $F(T)$ exhibits thermal duality order by order
in string theory.  Our conjecture does not alter this behavior.
What we are conjecturing, however, is that the sum of these
order-by-order perturbative functions $F(T)$ actually
exhibits an additional symmetry, one which guarantees
that the entropy $S(T)$ is also duality covariant.   
Thus, while thermal duality is indeed preserved order by order
for the string free energy, we are conjecturing
that the entropy, which normally fails to exhibit this symmetry
at any order, actually will exhibit this symmetry when all of these
separate order-by-order contributions are summed together.

Of course, this conjecture requires not only a special
temperature behavior at each order, but also a specific value of
the string coupling $\kappa$.  To see this, recall that 
the full free energy $F(\kappa,T)$ depends not only on the 
temperature $T$ but also on the string coupling $\kappa$.
Specifically, if $F_g(T)$ is the genus-$g$ contribution to the 
total free energy $F(\kappa,T)$, then 
\beq
          F(\kappa,T) ~=~ \sum_{g=1}^\infty \,\kappa^{2(g-1)} F_g(T)~.
\label{pertsum}
\eeq
In general, the genus-$g$ free energy transforms as a
weight-$2g$ duality-covariant function,
\beq
         F_g(T_c^2/T) ~=~ (T_c/T)^{2g} \, F_g(T)~,
\eeq
which is why the total free energy, like its genus-one
contribution, transforms
as a weight-two duality-covariant function:  
\beq
        F(\kappa,T_c^2/T) ~=~ (T_c/T)^2 \,F(\kappa T_c/T, T)~.
\eeq
The corresponding shift in the string coupling is precisely analogous 
to what occurs in T-duality.
However, since the string coupling $\kappa$ parametrizes the relative
weightings of the contributions from each genus, 
any new symmetry 
which appears only in the sum over all genera 
must hold only for a specific value of the string coupling. 
Our conjecture, which claims that the full free energy $F(\kappa,T)$ 
must have the exact temperature dependence given by $F^{(\ell)}(T)$
with $\ell=D$,
must therefore hold only for a specific value of
the string coupling which in turn must presumably be fixed by
other, non-perturbative effects.

While these are exciting speculations,
we are nevertheless left with our original question
as to whether there exist special finite-temperature string ground
states for which {\it all}\/ relevant thermodynamic 
quantities exhibit thermal duality.
For $D>2$, it seems that such states do {\it not}\/ exist:
even if the above conjecture is correct and the exact effective
potentials of such string models match our 
$\ell>2$ functional forms,
these functional forms do not preserve thermal duality
covariance for the specific heat.
Only the $\ell=1,2$ solutions have this property. 
However, for $D=2$, the answer to this question
may be somewhat more positive, for the case of 
time/temperature circle compactifications with $Z_{\rm model}=1$ 
appears to yield exactly what we require.  Thus, even when we take 
$Z_{\rm model}\not = 1$,
our above conjecture suggests that  
the corrections that are induced 
by the non-trivial $Z_{\rm model}$ 
might ultimately disappear when contributions from all
orders are included.  
Indeed, in this way, our conjecture would lead to a model-independent 
universal form for the effective potentials corresponding to
such compactifications. 
However, it is important to realize that even if the circle case 
leads to a duality-covariant entropy and specific heat,
the corresponding orbifold case certainly does not.
Since the additive shift in the effective potential that accrues in passing from
the circle to the orbifold is given by $3/2$ rather than $\delta= \pm 1$,
the orbifold case corresponds not to the $\ell=1$ solution but rather
to a {\it linear combination}\/ of the $\ell=1$ and $\ell=2$ solutions,
as indicated in Eq.~(\ref{lincomb}).
The resulting entropy is thus a linear combination of two
terms with different duality weights, and 
fails to be covariant at all.
Of course, the specific heat continues to be covariant,
since the specific heat is unaffected by the contributions
from the orbifold fixed points.

What then are we to conclude from this analysis?
Clearly, if string theory is to resurrect thermal duality
for quantities such as entropy and specific heat, 
the miracle
is not likely to lie in the clever choice of a string ground state.
Rather,
the miracle is more likely to lie in the structure
of thermodynamics itself, as a possible string-theoretic modification 
of the usual rules of classical thermodynamics according
to which quantities such as entropy and specific heat are calculated. 
Indeed, as we shall see in Ref.~\cite{II},
such an approach is capable of restoring thermal duality
to {\it all}\/ thermodynamic quantities --- regardless of the specific ground state ---
and leads to a new, manifestly duality-covariant string thermodynamics.
The development of such a theory
will be explored in Ref.~\cite{II}.

\section*{Acknowledgments}
\setcounter{footnote}{0}

This work is supported in part by the National Science Foundation
under Grants~PHY-0071054 and PHY-0301998,
and by a Research Innovation Award from Research Corporation.
We wish to thank D.~Marolf, R.~Myers, R.~Roiban,
C.~Stafford, and G.~Torrieri for discussions.

\bigskip
\vfill\eject

\bibliographystyle{unsrt}

\end{document}